\pdfoutput=1
\documentclass[aps,prd,reprint,nofootinbib,superscriptaddress,preprintnumbers,showkeys]{revtex4-1}

\usepackage{color} 
\usepackage{amsmath} 
\usepackage{amssymb} 
\usepackage{txfonts} 
\usepackage{paralist} 
\usepackage{verbatim} 
\usepackage{graphicx} 
\usepackage{xspace} 
\usepackage{hyperref} 
\usepackage{comment}
\usepackage[capitalise]{cleveref}

\usepackage[acronym,shortcuts]{glossaries}
\glsdisablehyper

\newacronym[user1={gravitational-wave}]{GW}{GW}{gravitational wave}
\newacronym{GRB}{GRB}{gamma-ray burst}
\newacronym{SGRB}{SGRB}{short-duration gamma-ray burst}
\newacronym{LHO}{LHO}{\protect\ac{LIGO} Hanford Observatory}
\newacronym{LIGO}{LIGO}{Laser Interferometer Gravitational Wave Observatory}
\newacronym{LLO}{LLO}{\protect\ac{LIGO} Livingston Observatory}
\newacronym{S5VSR}{S5/VSR1}{\protect\ac{LIGO} Science Run 5 and Virgo Science Run 1}
\newacronym{S6VSR}{S6/VSR2/VSR3}{\protect\ac{LIGO} Science Run 6 and Virgo Science Runs 2 \& 3}
\newacronym{SNR}{SNR}{signal-to-noise ratio}
\newacronym[\glslongpluralkey={false alarm probabilities}]{FAP}{FAP}{false alarm probability}
\newacronym{PSD}{PSD}{power spectral density}
\newacronym{FFT}{FFT}{fast Fourier transform}
\newacronym{CBC}{CBC}{compact binary coalescence}
\newacronym{BAT}{BAT}{Burst Alert Telescope}
\newacronym{GBM}{GBM}{Gamma-ray Burst Monitor}
\newacronym{IPN}{IPN}{InterPlanetary Network}
\newacronym{GCN}{GCN}{Gamma-ray Coordination Network}
\newacronym{NSBH}{NSBH}{neutron star--black hole}
\newacronym{BNS}{BNS}{binary neutron star}

\newglossaryentry{ltt}{
    name={light travel time},%
    description={time taken for a light beam to travel directly between %
                 two detectors}}

\newcommand{\degree}{\ensuremath{{}^{\circ}}\xspace} 
\newcommand{\dccversion}{LIGO-P1400044-v4} 
\newcommand{\HF}{Harry:2010fr}

\begin{document}

\title{Improved methods for detecting gravitational waves associated with short gamma-ray bursts}
\author{A. R. Williamson}
\affiliation{School of Physics and Astronomy, Cardiff University, Cardiff, UK}
\author{C. Biwer}
\affiliation{Department of Physics, Syracuse University, Syracuse, NY 13244, USA}
\author{S. Fairhurst}
\affiliation{School of Physics and Astronomy, Cardiff University, Cardiff, UK}
\author{I. W. Harry}
\affiliation{School of Physics and Astronomy, Cardiff University, Cardiff, UK}
\affiliation{Department of Physics, Syracuse University, Syracuse, NY 13244, USA}
\author{E. Macdonald}
\affiliation{School of Physics and Astronomy, Cardiff University, Cardiff, UK}
\author{D. Macleod}
\affiliation{School of Physics and Astronomy, Cardiff University, Cardiff, UK}
\affiliation{Louisiana State University, Baton Rouge, LA 70803, USA}
\author{V. Predoi}
\affiliation{School of Physics and Astronomy, Cardiff University, Cardiff, UK}
\date{\today} 
\preprint{\dccversion}

\begin{abstract}
In the era of second generation ground-based gravitational wave detectors, short gamma-ray bursts (GRBs) will be among the most promising astrophysical events for joint electromagnetic and gravitational wave observation.
A targeted search for gravitational wave compact binary merger signals in coincidence with short GRBs was developed and used to analyze data from the first generation LIGO and Virgo instruments.
In this paper, we present improvements to this search that enhance our ability to detect gravitational wave counterparts to short GRBs.
Specifically, we introduce an improved method for estimating the gravitational wave background to obtain the event significance required to make detections; implement a method of tiling extended sky regions, as required when searching for signals associated to poorly localized GRBs from \textit{Fermi} Gamma-ray Burst Monitor or the InterPlanetary Network; and incorporate astrophysical knowledge about the beaming of GRB emission to restrict the search parameter space.
We describe the implementation of these enhancements and demonstrate how they improve the ability to observe binary merger gravitational wave signals associated with short GRBs.  
\end{abstract}

\maketitle

\section{Introduction}
\label{sec:intro}
\Acp{GRB} are amongst the most energetic electromagnetic events in the universe, observed isotropically across the sky and up to cosmological redshifts~\cite{Paciesas:1999tp}.
An apparent bimodality observed in the duration and spectral hardness of GRBs -- long-soft and short-hard --- suggests more than one class of progenitors~\cite{Kouveliotou:1993yx}.  
The mergers of compact binary systems composed of two neutron stars or a neutron star and a black hole have long been proposed as possible progenitors of short \acp{GRB}~\cite{Eichler:1989ve,Narayan:1992iy}.
Short \ac{GRB} variability timescales are small, indicating a compact source~\cite{Rees:1994nw}, while the observed offsets from their host galaxies agree with that expected for a population of compact binary mergers and not of core-collapse supernovae~\cite{Church:2011gk}, the widely accepted progenitors of most long \acp{GRB}.
The recent detection of a kilonova associated with GRB130603B~\cite{Berger:2013wna,Tanvir:2013pia} has further supported the compact merger hypothesis.
For an in-depth review of short \ac{GRB} science, see eg.~\cite{Nakar:2007yr,Berger:2013jza}.

If short \acp{GRB} are indeed compact binary mergers, they are a very interesting class of events for \ac{GW} astronomy, since such compact binary mergers are also strong emitters of \acp{GW}~\cite{Cutler:1992tc,Shibata:2007zm}.
\ac{GW} observations of \acp{GRB} will make possible direct observation of the central engines that power these events, a feat that electromagnetic observations alone cannot achieve due to circumburst material and ejecta~\cite{Bartos:2012vd}.
The observation of a short \ac{GRB} provides the time and sky position of a potential \ac{GW} source.  
A targeted search for a binary merger \ac{GW} signal, informed by the \ac{GRB} observation, need only search a small fraction of the parameter space of an un-triggered, full-sky, binary merger search.  
Consequently, it is possible to significantly reduce the detection threshold for the targeted GRB search~\cite{Clark:2014ut}, thereby increasing the sensitivity of the search.

An analysis pipeline has been developed specifically for performing the targeted search for binary mergers associated with \acp{GRB} \cite{\HF}.
At its heart, this is a matched-filtering analysis~\cite{Wainstein:1962es} that makes use of the well understood gravitational waveforms emitted during binary merger to search for a signal in data from the operational \ac{GW} detectors.
The analysis makes use of the known sky location of the \ac{GRB} and the relative \ac{GW} detector sensitivities to appropriately time shift and weight the data streams from the individual detectors to perform a \textit{coherent} analysis.
This is in contrast to a \textit{coincidence} analysis, which individually analyzes the data from the different detectors and then performs a coincidence check (see e.g.~\cite{Babak:2012zx}).  
By coherently combining the data, it is possible to isolate data streams containing the two gravitational wave polarizations.
In the case where data from more than two \ac{GW} detectors are used, the other, orthogonal contributions provide a \textit{null stream} which helps to reduce the impact of non-stationary noise.  
The analysis pipeline also makes use of signal consistency tests to check that anything causing a large \ac{SNR} is consistent with a putative \ac{GW} signal.
As well as using these tests to reject noise transients, or \textit{glitches}, they are also used to down-weight events which are more consistent with a noise transient than a signal.
The significance of the results is computed by comparing the result for the six seconds around the time of the \ac{GRB} with data from surrounding times, thereby calculating the probability of obtaining an event with a specific \ac{SNR} due to noise alone.

The analysis pipeline described above has been used to carry out numerous \ac{GRB} searches on data from the initial \ac{LIGO}~\cite{Abbott:2007kv} and Virgo~\cite{Accadia:2012zzb} detectors.  
Searches associated with short \acp{GRB} observed by the \textit{Swift}~\cite{Barthelmy:2005hs} and \textit{Fermi}~\cite{Meegan:2009qu} satellites have been performed~\cite{Abadie:2010uf,Briggs:2012ce}, as well as those observed by the \ac{IPN}~\cite{Hurley:2002wv,Aasi:2014iia}.
None of these analyses made a \ac{GW} detection in conjunction with an observed short \ac{GRB}, which was not surprising given the sensitivity of the initial detectors --- tens of Mpc for binary merger signals --- and the typical distances to \acp{GRB} --- a median redshift of 0.7 and a closest measured redshift of 0.1, implying a distance of 500 Mpc.
The second generation of \ac{GW} detectors, Advanced LIGO~\cite{Harry:2010zz} and Advanced Virgo~\cite{AdVirgo}, are due to begin observing in 2015 with sensitivities that will increase, over time, to approximately ten times greater than those of the first generation detectors \cite{Aasi:2013wya}.
The prospects for detection of \ac{GW} signals associated with \acp{GRB} with the advanced detectors are promising~\cite{Clark:2014ut,Abadie:2010cf,Dietz:2012qp,Chen:2012qh}.

With the realistic prospect of a joint \ac{GW}--\ac{GRB} observation in the coming years, we have made a number of changes and improvements to the analysis pipeline.  
These enhancements are critical to optimizing the potential for observing binary merger \ac{GW} signals in coincidence with electromagnetically observed short \acp{GRB}.
Most importantly, we have improved the ability of the pipeline to estimate the significance of rare events.
To this end, we have introduced the ability to perform time-shifted analyses, whereby the data from the different detectors are shifted by several seconds relative to each other and the analysis is repeated.  
This allows us to measure the background of the search to lower than one part in $10^{5}$, a level that would be required for an unambiguous detection claim \cite{Colaboration:2011np}.

The targeted search introduced in~\cite{\HF} makes use of the sky location of the source, but places no restrictions on the orientation of the binary.  
Observations of short \acp{GRB} indicate that the gamma-ray jet is beamed, with most observations favoring a beaming angle of $30 \degree$ or smaller~\cite{Guelbenzu:2012id}, with the GRB jet emitted orthogonal to the orbital plane of the binary.  
Thus, it is natural to incorporate this into the search by restricting the search to binaries which are observed to be (close to) face-on. 
This restriction reduces the parameter space of the search, providing an increase in sensitivity.

While \textit{Swift} provides typical \ac{GRB} localizations with arcminute accuracy \cite{Barthelmy:2005hs}, \textit{Fermi}'s \ac{GBM} and the \ac{IPN} often provide localizations to significantly larger regions of the sky ~\cite{Meegan:2009qu,Hurley:2002wv}.  
For these \acp{GRB}, the $3 \sigma$ confidence sky localization region can be tens of square degrees so it no longer suffices to search a single sky point.  
We describe a method of searching over a grid of sky points that cover the uncertainty region, and demonstrate that by using this grid we can efficiently search for a signal originating from any point in the sky patch.
This method has already been used in the searches described in~\cite{Briggs:2012ce,Aasi:2014iia}.

The layout of this paper is as follows.
In \cref{sec:summary} we review the coherent analysis pipeline introduced in~\cite{\HF}.
In \cref{sec:back} we describe and demonstrate the improved background calculation. 
In \cref{sec:face} we describe how searching only for signals with a narrow opening angle can improve search sensitivity.
In \cref{sec:sky} we describe how to tile search points on the sky and explore how this improves sensitivity for GRBs with larger sky localization regions.
Finally, in \cref{sec:discuss} we discuss the implications of our results.

\section{Pipeline Summary}
\label{sec:summary}
The targeted, coherent search was described in detail in~\cite{\HF}.
Here, we provide a brief review of the analysis pipeline in order to provide the necessary background for the following sections.
Where possible, we follow the same notation as the original paper.
For a more detailed description see~\cite{\HF} itself.

The targeted, coherent search is carried out whenever an observed \ac{GRB} is detected during a time that at least two gravitational wave detectors are operating and have good quality data for a sufficiently long period of time either side of the \ac{GRB}.
In practice, we search for a signal in a 6 second window covering 5 seconds before to 1 second after the Earth crossing time of the \ac{GRB} called the \textit{on-source} window.
However, we require additional data around this time in order to perform the analysis to ensure that the detectors were operating stably at the time of the \ac{GRB} and to provide a good estimate of the detector sensitivity at the time.
Our ability to detect a \ac{GW} signal associated with a \ac{GRB} depends upon both the stationary noise background and also the non-stationary noise transients in the data which might mask a signal.  
The data surrounding the on-source time is used to evaluate both of these.

\subsection{Multi-Detector Matched Filter}

The pipeline performs a modelled gravitational wave search for compact binary inspiral signals.
To this end, a bank of template waveforms~\cite{Blanchet:2013haa,Owen:1998dk} that densely cover the mass parameter space is used to perform a matched-filter analysis~\cite{Wainstein:1962es}.
The targeted \ac{GRB} search performs a coherent analysis whereby the data streams from different detectors are combined while performing the matched-filtering and a network \ac{SNR} is calculated directly.
The majority of searches for binary merger \acp{GW} perform the matched-filter independently on individual interferometer data streams before comparing the resulting \textit{triggers} to search for coincident events (see e.g.~\cite{Babak:2012zx}).
The coherent analysis affords several benefits.
First, by performing the analysis coherently, we combine the detector data to produce two data streams which are sensitive to the two gravitational wave polarizations.
Any other, orthogonal data streams will necessarily contain only noise and can either be ignored, or used to eliminate noise transients which will often contribute power to these \textit{null streams}.
Additionally, by combining the data from the detectors at the time of analysis, we will accumulate power from all detectors, not just those which produced a trigger above threshold.
It was shown in~\cite{\HF} that a coherent analysis provides an improvement in sensitivity over the coincident one, but the search is more computationally costly than a coincident one.
A targeted \ac{GRB} search, where both the sky location and arrival time of the signal are constrained is ideal for performing the more sensitive, coherent analysis.

The amplitude of a \ac{GW} signal from a non-precessing binary may be decomposed into two polarizations, denoted $+$ and $\times$, as
\begin{subequations}\label{eq:ht}
    \begin{align}
        h_+(t) &= \mathcal{A}^1h_0(t) + \mathcal{A}^3h_{\pi/2}(t) \, , \\
        h_\times(t) &= \mathcal{A}^2h_0(t) + \mathcal{A}^4h_{\pi/2}(t) \, .
    \end{align}
\end{subequations}
Here, $h_{0}$ and $h_{\pi/2}$ denote the two phases of the waveform, which depend upon the binary masses as well as the coalescence time of the signal.
These are calculated using the post-Newtonian formalism~\cite{Blanchet:2013haa}.
In the analysis presented here, we restrict to non-spinning components.
However, the search is easily extended to binaries with spins aligned with the orbital angular momentum by simply generating additional templates to cover the spin parameter space (see e.g.~\cite{Harry:2013tca,Brown:2012qf}).
The amplitude terms for an inspiral \ac{GW} signal are
\begin{subequations}
    \label{eq:amps}
    \begin{align}
        \mathcal{A}^1 &= \frac{D_0}{D} \frac{\left(1 + \cos^2 \iota \right)}{2} \cos 2 \phi_0 \cos 2 \psi  \nonumber\\* & \qquad {} 
        - \frac{D_0}{D} \cos \iota \sin 2 \phi_0 \sin 2 \psi  \, , \\
        \mathcal{A}^2 &= \frac{D_0}{D} \frac{\left(1 + \cos^2 \iota \right)}{2} \cos 2 \phi_0 \sin 2 \psi  \nonumber\\* & \qquad {} 
        + \frac{D_0}{D} \cos \iota \sin 2 \phi_0 \cos 2 \psi  \, , \\
        \mathcal{A}^3 &= -\frac{D_0}{D} \frac{\left(1 + \cos^2 \iota \right)}{2} \sin 2 \phi_0 \cos 2 \psi  \nonumber\\* & \qquad {} 
        - \frac{D_0}{D} \cos \iota \cos 2 \phi_0 \sin 2 \psi  \, , \\
        \mathcal{A}^4 &= -\frac{D_0}{D} \frac{\left(1 + \cos^2 \iota \right)}{2} \sin 2 \phi_0 \sin 2 \psi  \nonumber\\* & \qquad {} 
        + \frac{D_0}{D} \cos \iota \cos 2 \phi_0 \cos 2 \psi  \, .
    \end{align}
\end{subequations}
These terms are dependent on 4 variables: the source distance, $D$; the coalescence phase, $\phi_0$; the polarization angle, $\psi$; and the inclination angle, $\iota$.
$D_0$ is a scaling distance (usually 1 Mpc).
It is worth noting that, for any set of amplitudes $\mathcal{A}^{\mu}$, there is a unique set of $\{D, \iota, \phi_0, \psi\}$, up to reflection and rotation symmetry.

The \ac{GW} signal seen by a detector $X$ is a combination of the two polarizations, each weighted by an antenna power pattern factor $F_{\{+,\times\}}$ \cite{Jaranowski:1998qm}, which describes the relative response of the detector to each polarization,
\begin{equation}
    h^X(t) = F_+^X h_+(t^{X}) + F_\times^X h_\times(t^{X}) \, .
\end{equation}
Here, $t^{X}$ is the time of arrival of the signal at detector $X$, which will depend upon a fiducial arrival time (for example at the geocenter) and the relative location of the detector and source.

In matched-filtering analysis the inner products between a template gravitational waveform time series $h(t)$ and detector data stream time series $s(t)$ are calculated.
In general, the inner product between two such time series, $a^X$ and $b^X$, is given by
\begin{equation}
    \left(a^X \middle| b^X\right) = 4 \operatorname{Re} \int\limits_0^\infty \frac{\tilde{a}^X(f) \cdot \tilde{b}^X(f)^*}{S_h^X(f)} \, ,
\end{equation}
where $S_h^X(f)$ is the noise power spectral density in detector $X$, and $\tilde{a}(f)$ denotes the Fourier transform of the time series $a(t)$.
For binary merger signals, the two phases $h_{0}$ and $h_{\pi/2}$ are orthogonal, in the sense that
\begin{equation}
    (h_{0} | h_{\pi/2} )  = 0 \, .
\end{equation}

For a network of detectors, we define the multi-detector inner product as the sum of the single detector inner products,
\begin{equation}
    (\mathbf{a | b}) \equiv \sum_{X=1}^{d} \left(a^X \middle| b^X\right) \, ,
\end{equation}
where $d$ denotes the number of detectors in the network.
The multi-detector log-likelihood is then defined as,
\begin{align}
    \label{eq:loglike}
    \ln \Lambda &= (\mathbf{s | h}) - \frac{1}{2} (\mathbf{h | h}) \nonumber \\
                &= \left[ \mathcal{A}^\mu (\mathbf{s|h_\mu}) - \frac{1}{2}\mathcal{A}^\mu \mathcal{M}_{\mu \nu} \mathcal{A}^\nu \right] \, ,
\end{align}
where $\mathbf{h} = (\mathbf{F_+ h_0, F_\times h_0, F_+ h_{\pi/2}, F_\times h_{\pi/2}})$, and the matrix
\begin{equation}
    \mathcal{M}_{\mu\nu} \equiv (\mathbf{h_\mu | h_\nu}) \, .
\end{equation}
Maximizing this likelihood ratio over the amplitude parameters $\mathcal{A}_{\mu}$, we obtain the maximized coherent \ac{SNR},
\begin{equation}\label{eq:coh_snr}
    \rho^2_{\mathrm{coh}} \equiv 2 \ln \Lambda |_{\mathrm{max}} = \left[ (\mathbf{s | h_\mu}) \mathcal{M}^{\mu\nu} (\mathbf{s | h_\nu}) \right] \, ,
\end{equation}
where $\mathcal{M}^{\mu\nu}$ is the inverse of the matrix $\mathcal{M}_{\mu\nu}$.

The coherent \ac{SNR} forms the basis of the detection statistic and has a $\chi^2$ background distribution with 4 degrees of freedom.
The four degrees of freedom correspond to the four components of the gravitational wave signal -- the $0$ and $\pi/2$ phases of the two polarizations.
This becomes more transparent if we work in the dominant polarization frame.  
In this frame, the network is maximally sensitive to the $+$ polarization and the two polarizations are orthogonal.
Then, the coherent SNR can be re-expressed as
\begin{equation}
    \rho^2_{\mathrm{coh}} = \frac{\mathbf{(s | F_{+} h_{0})}^{2} + \mathbf{(s | F_{+} h_{\pi/2})}^{2}}{\mathbf{(F_{+} h_{0} | F_{+} h_{0})}} +
    \frac{\mathbf{(s | F_{\times} h_{0})}^{2} + \mathbf{(s | F_{\times} h_{\pi/2})}^{2}}{\mathbf{(F_{\times} h_{0} | F_{\times} h_{0})}} \, .
\end{equation}

In Gaussian noise, the coherent \ac{SNR} would be the detection statistic.
Events with a larger coherent \ac{SNR} would be less likely to be due to noise fluctuations and consequently more likely to be due to a \ac{GW} signal.
However, in real data \ac{GW} signals are not the only cause of deviations from the background distribution.
Noise transients, or \textit{glitches}, also contribute to the background.
Although glitches will not typically mimic template waveforms, if they are large enough they will still produce a large \ac{SNR}.
Consequently, we must use a number of consistency tests to eliminate or down-weight triggers that are unlikely to be due to a \ac{GW} signal incident upon the detector network.

\subsection{Signal Consistency}

Matched filtering alone leads to the identification of a large number of triggers, many of which are purely due to non-Gaussian noise transients present in the data streams.
Such noise transients may be discarded by performing signal consistency tests across the individual detectors that make up the network.
Here, we briefly describe the different tests used in the analysis.

\subsubsection{Null Stream Consistency}

Null stream consistency makes use of one or more null data streams or, in the case of this pipeline, the related null \ac{SNR} statistic.
This is simply the \ac{SNR} observed in the detector network that is \textit{not} consistent with the signal model:
\begin{equation}
\rho_{N}^{2} \equiv \sum_{X} \rho_{X}^{2} - \rho_{\mathrm{coh}}^{2} \, ,
\end{equation}
where $\rho_X$ is the \ac{SNR} in detector $X$. For a signal which matches the template waveform, there will be no signal power in the null \ac{SNR}, and it will be $\chi^2$ distributed with $2d - 4$ degrees of freedom due to the presence of noise.
An incoherent, non-Gaussian transient noise event will contribute to the null \ac{SNR} and consequently a large null \ac{SNR} is used to eliminate spurious events via a hard cut if
\begin{equation}
    \label{eq:null_cut}
    \begin{aligned}
        \rho_{\mathrm{null}} &> 5.25, &&\rho_{\mathrm{coh}} \leq 20 \\
        \rho_{\mathrm{null}} &> \frac{\rho_{\mathrm{coh}}}{5} + 5.25, &&\rho_{\mathrm{coh}} > 20
    \end{aligned} \, .
\end{equation}

\subsubsection{Single detector thresholds}

Noise transients are, by their nature, events which occur in a single detector.
Conversely, gravitational wave events will lead to signal power being distributed among all detectors in the network.
We can use this difference to further reduce the background due to glitches.
The most effective, and most straightforward, method is simply to require that a signal is observed with an \ac{SNR} above threshold (typically four) in at least two detectors.
This serves to eliminate the majority of glitches, which have power in only one detector, with very little effect on signals.

\subsubsection{$\chi^2$ Tests}

When matched-filtering identifies a trigger with a large \ac{SNR} there is necessarily some component of the data which matches the signal $h(t)$.
If the trigger is caused by a noise glitch, there is likely to be an additional, orthogonal component of the data which is not well described by Gaussian noise.
$\chi^2$ tests are designed to eliminate glitch triggers by identifying power that is not consistent with either signal or Gaussian noise.
To do so, we introduce a set of basis waveforms $T^{i}$ which are orthonormal and also orthogonal to the signal waveform $h(t)$.
Specifically, we require
\begin{equation}
    \mathbf{(T^{i}_{\mu} | T^{j}_{\nu})} = \delta^{ij}\delta_{\mu\nu} 
    \quad \mbox{and} \quad
    \mathbf{(T^{i}_{\mu} | h_{\nu})} = 0 \, ,
\end{equation}
where $\mu, \nu$ refer to the waveform components and $i,j$ the waveforms that comprise the basis for the $\chi^{2}$ test.
We then construct a $\chi^{2}$ statistic as
\begin{equation}
    \chi^2 = \sum_{\mu=1}^4 \sum_{i=1}^N (\mathbf{T}_\mu^i | \mathbf{s})^2 \, .
\end{equation}
In the presence of a signal that matches the template waveform (or no signal), the statistic will be $\chi^{2}$ distributed with 4N degrees of freedom.
If the data contains some additional, non-Gaussian noise the $\chi^{2}$ value will be elevated provided that the set of templates $T^{i}$ captures at least a fraction of the power contained in the glitch.
Triggers with a large $\chi^{2}$ value are discarded.
In practice it is far from trivial to choose the set of waveforms $T^i$ so that they are both orthonormal and orthogonal to $h(t)$, and match a variety of non-Gaussianities.
Three different $\chi^{2}$ tests have been implemented in the analysis:

\begin{enumerate}[i]
\item \textit{Frequency bins:}
    The test waveforms $T^{i}$ are generated by chopping up the template $h(t)$ into (N+1) sub-templates in the frequency domain, each of which contains an equal amount of power.
    From these, we generate N orthonormal waveforms which are also orthogonal to $h(t)$.
\item \textit{Template bank:}
    The test waveforms $T^{i}$ are taken from the template-bank of binary merger waveforms used in the search.
    In general, these will not be orthogonal to $h(t)$, but it is straightforward to subtract the part proportional to $h(t)$.
    However, it is more difficult to render the waveforms $T^{i}$ orthonormal.
    In practice we do not attempt to do so, but instead use an empirical threshold based on an effective number of degrees of freedom.
\item \textit{Autocorrelation:}
    The test waveforms $T^{i}$ are simply copies of the waveform $h(t)$ offset in time from the original.
    As with the template bank, it is straightforward to remove the component of $T^{i}$ that is proportional to $h(t)$.
    We do not attempt to orthonormalize the $T^{i}$ and again empirically set the threshold.
\end{enumerate}

\subsubsection{Re-weighted SNR}

In addition to discarding triggers which fail the signal consistency test described above, we also re-weight the \ac{SNR} of triggers based on the values of the $\chi^{2}$ tests and null \ac{SNR}.
This allows us to better differentiate signals from noise background.
The re-weighting is chosen such that the \ac{SNR} of signals will be unaffected while those noise triggers which do not match well with the template waveform will be down-weighted.
We perform two sets of down-weighting.
Firstly, with the frequency bin $\chi^{2}$ values,
\begin{equation}
    \label{eq:snr_chisq}
    \rho_{\chi^2} =
    \begin{cases}
        \rho_{\mathrm{coh}} & \chi^2 \leq n_{\mathrm{dof}} \\
        \frac{\rho_{\mathrm{coh}}}{\left\{ \left[ 1+ \left( \frac{\chi^2}{n_{\mathrm{dof}}} \right)^3 \right] / 2 \right\}^{1/6}} & \chi^2 > n_{\mathrm{dof}}
    \end{cases} \, ,
\end{equation}
and then the null \ac{SNR},
\begin{equation}
    \rho_{\mathrm{rw}} =
    \begin{cases}
        \rho_{\chi^2} & \rho_{\mathrm{null}} \leq 4.25 \\
        \frac{\rho_{\chi^2}}{\rho_{\mathrm{null}} - 3.25} & \rho_{\mathrm{null}} > 4.25
    \end{cases} \, .
\end{equation}
This re-weighted \ac{SNR} value is the detection statistic used for evaluating candidate events.

We note that the $\chi^{2}$ re-weighted \ac{SNR} given in \cref{eq:snr_chisq} is different from the one used in the original paper~\cite{\HF}.
In particular, the exponents in the denominator have been changed.
In the process of developing an all-sky, all-time coherent analysis~\cite{all_sky_coh}, it was found that the original re-weighting left a small tail of high \ac{SNR} noise events.
These had not been observed in the \ac{GRB} search previously, due to the limited amount of data used in the analyses.
By using a re-weighted \ac{SNR} identical to the one used in the all-sky coincidence search \cite{Babak:2012zx}, we were able to eliminate the high SNR events.
The same re-weighting has now been applied in the \ac{GRB} search.

\subsection{Event significance}

The analysis described above is performed for all template waveforms in the template bank covering the mass space.
For each template, the re-weighted \ac{SNR} is calculated.
The template producing the largest re-weighted \ac{SNR} during the on-source window is retained as the event candidate.

The significance of this event is calculated using the data before and after the time of the \ac{GRB}, which is designated \textit{off-source}.
This data will not contain a signal corresponding to the \ac{GRB} and is also unlikely to contain a \ac{GW} signal from the same sky position which is unassociated with the \ac{GRB}, thus any events occurring in the off-source will be due to background noise.
In a typical search we use approximately an hour of data for the off-source, and split this into trials with durations equal to that of the on-source window.
This gives us a means of characterizing the background noise in our detector network around the time of the \ac{GRB}.
The significance of the on-source event is determined by calculating the false alarm probability, or p-value.
This is simply the fraction of off-source trials with an event of equal or greater significance than in the on-source.

To evaluate the sensitivity of the pipeline to finding \ac{GW} signals in the data around the time of the \ac{GRB}, we inject a number of simulated signals into the off-source data.
The simulated signals are drawn randomly from an astrophysically motivated distribution of distances, component masses and spins and binary inclination.
The simulated signals are compact binary merger waveforms at $3.5$ post-Newtonian order~\cite{Blanchet:2004ek,Blanchet:2013haa}, where one component of the binary is taken to be a neutron star and the second either a neutron star or black hole.
The efficiency of the analysis at recovering these signals provides a measure of pipeline performance and produces a lower limit on the distance to which the pipeline is sensitive.

\subsection{Example GRB}
In the remainder of this paper, we will illustrate the various pipeline developments using example analyses based upon the \ac{GRB} 100928A, which was observed by the \textit{Swift} \ac{BAT}~\cite{2010GCN..11310...1D,2010GCN..11312...1K}.
No other \textit{Swift} instrument observed this \ac{GRB} as the spacecraft was unable to slew to the sky position of the prompt burst due to a Sun observing constraint.
It was not detected by \textit{Fermi} or any other gamma ray sensitive instrument.

We have chosen this \ac{GRB} for a number of reasons.
Virgo and both \ac{LIGO} detectors were operational and had ample science quality data either side of the \ac{GRB} time.
Specifically, 5264 seconds of coherent network data between 01:34:35 and 03:02:19 UTC on 28 September 2010 was available for analysis purposes.
Additionally, the \ac{BAT} localized the burst to a point on the sky (RA $= 223.037\degree$, Dec $= -28.542\degree$) where both \ac{LIGO} detectors were approximately equally sensitive, and where Virgo had good sensitivity.
Furthermore, this position was known accurately, with a 90\% confidence radius of only 2.3 arcminutes.

\section{Background Estimation}
\label{sec:back}
To make a confident detection statement, we must establish that the probability of an observed event being due to noise alone is very small.
This requires a detailed understanding of the search background generated by both Gaussian detector noise and non-stationary transients.
We do this by looking at the data around the time of the \ac{GRB}.
We make the reasonable assumption that the off-source data contains no \ac{GW} signal originating from the same location on the sky and has, on average, the same statistical properties as the detector network background during the on-source period.
Thus, the off-source data provides a means of characterizing the background noise in the detector network at the time of the \ac{GRB}.

The \ac{FAP} associated to the on-source event, with re-weighted \ac{SNR} $\rho^{\star}$, is the probability of having a more significant event in any randomly chosen 6 seconds of data.
This is calculated by counting the fraction of background trials which have an event with $\rho > \rho^{\star}$,
\begin{equation}
    \text{FAP} = \frac{N\left(\rho > \rho^{\star} \right)}{N_{\mathrm{BG}}} \, ,
\end{equation}
where $N_{\mathrm{BG}}$ denotes the total number of background trials.
In the standard approach, we simply split the background into as many 6 second trials as possible, so the number of background trials is given by $N_{\mathrm{BG}} = T_{\mathrm{off}}/T_{\mathrm{on}}$.

The standard analysis makes use of approximately an hour of data around the time of the \ac{GRB}, leading to a lower limit on the \ac{FAP} of around $10^{-3}$.
For the majority of \acp{GRB}, this will be sufficient to demonstrate that there is no candidate \ac{GW} event associated to a particular \ac{GRB}.
However, when there is an interesting candidate, a \ac{FAP} of $10^{-3}$ is not sufficient to warrant a detection claim, and further background trials are required to more accurately evaluate the significance.

What would be an acceptable \ac{FAP} to support a detection claim?
In particle physics, the standard level is a ``5-sigma'' observation, or 1 in 3 million.
In a recent gravitational wave search~\cite{Colaboration:2011np}, a simulated signal was added to the data and recovered with a false alarm rate of 1 in 7,000 years, which was deemed sufficient to claim evidence for a detection.
Translating this to a \ac{GRB} search equates to a \ac{FAP} of $3 \times 10^{-6}$ for one of the 50 short \acp{GRB} observed each year.
Alternatively, we might consider the chance of there being an observable signal around the time of a \ac{GRB}.
In~\cite{Clark:2014ut}, this was estimated to be around 1\% for the advanced detector network operating at design.
Clearly, a detection candidate would require a \ac{FAP} much lower than the probability of observing a signal.
All arguments point to requiring a minimum of $10^{5}$ background trials to assess the significance of a detection candidate.

To reach a significance level of $10^{-5}$, we require further background trials.
The most straightforward approach would be to simply extend the off-source analysis to incorporate one week of data.
While in principle this is possible, the typical duration of continuous operation for the detectors is on the order of hours.
Furthermore, the data quality is known to change between different stretches of data, so a week of off-source data may not accurately characterize the data at the time of the \ac{GRB}.
In addition, extending the off-source data to one week would increase the computational cost of the analysis by a factor of several hundred, rendering it impractical to estimate the background promptly.
Consequently, an alternative method is required.
To obtain an improved estimate of the network background, we instead artificially time shift the data from the different detectors and repeat the analysis.
These time shifts are always significantly longer than the light travel time between detectors and the signal auto-correlation time and typical glitch durations (all well under one second), so that \ac{GW} signals will not appear coherently in the time-shifted analysis.

\begin{figure}[tbp]
    \centering
    \includegraphics[width=\linewidth]{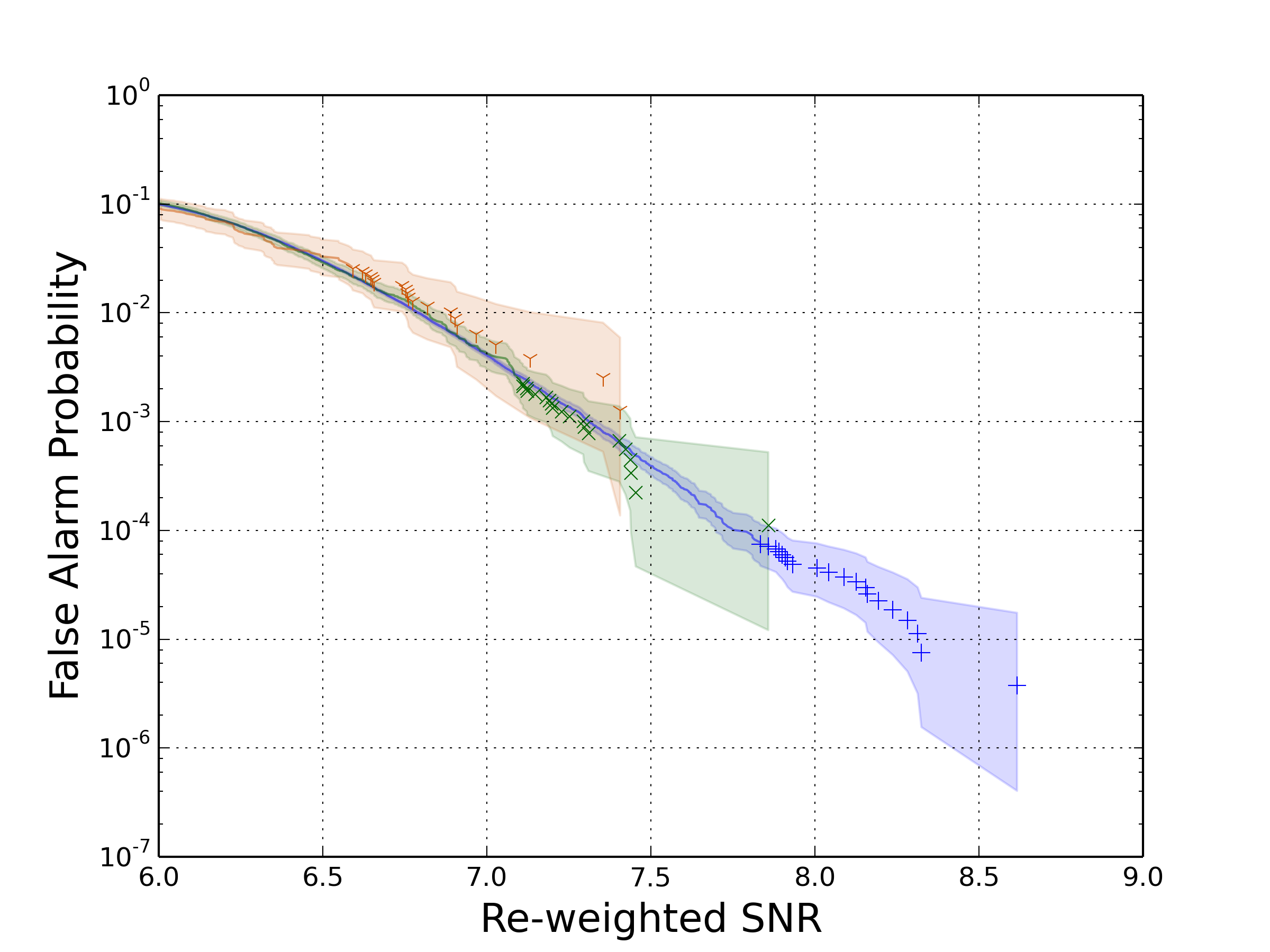}
    \caption{\Ac{FAP} as a function of the re-weighted \ac{SNR} detection statistic for a search performed for \ac{GRB} 100928A, using time slides to reach \ac{FAP} $< 10^{-5}$.
             The figure shows the background estimated with off-source only (787 trials) plotted in orange Y; the short slide analysis (8917 trials)  plotted in green $\times$; both long and short slides (267185 trials) plotted in blue +.
             With short slides alone, we can estimate a significance of 1 part in $10^{4}$ while long and short slides give a background estimate to 1 in $4\times 10^{6}$.
             The shaded regions show the 95\% Jeffreys credible interval for each case, which assumes each time slide is a statistically independent trial.
             For clarity of presentation we have only plotted the 20 loudest trials for each search.
    \label{fig:background_slides}
    }
\end{figure}

We are able to increase the number of background trials performed by an order of magnitude, with minimal impact on the computational cost, thereby allowing us to estimate \acp{FAP} to around $10^{-4}$.
This is achieved by time shifting the \ac{SNR} time series of the individual detectors prior to performing the coherent analysis.
In the analysis, the detector data is split into sections, typically of 128 seconds length, which are match filtered to produce a (complex) \ac{SNR} time series for each detector.
These are then combined according to \cref{eq:coh_snr} to calculate the coherent \ac{SNR} time series.
A \textit{short slide} is performed by introducing relative time-shift between the detectors' \ac{SNR} time series prior to computing the coherent \ac{SNR}.
For the example \ac{GRB}, we leave the H1 data alone, shift the L1 data by multiples of 6 seconds and the V1 data by multiples of 12 seconds.
This allows for ten time shifted analyses to be performed.
Since calculating the single detector \ac{SNR} time series is the most computationally costly part of the analysis, short slides have a relatively small computational cost.
In \cref{fig:background_slides}, we show the improvement in background estimation afforded by the inclusion of the short slides.

We have also implemented \textit{long slides} which involve permuting the data segments prior to analysis.
Unfortunately, this does require repeating the analysis, so the computational cost increases linearly with the number of long slides.
However, it is possible to perform short slides within each long slide.
Thus, we only require around ten long slides in order to achieve a background estimate of $10^{-5}$.

\cref{fig:background_slides} shows \ac{FAP} as a function of re-weighted \ac{SNR} for the analysis of \ac{GRB} 100928A.
This shows that any on-source event with $\rho_{\mathrm{rw}} > 8.5$ would have a \ac{FAP} at the $10^{-5}$ level.
We have, however, assumed that all time slides are independent.
In reality, all time slides are formed from different combinations of the same detector data streams, and so are not statistically independent at all.
A more rigorous treatment of \ac{FAP} uncertainty when dealing with time slides would likely show far larger 95\% credible intervals for all cases, however it is not clear how to implement such a treatment for this search~\cite{Was:2009vh}.

It is interesting to compare the background for the \ac{GRB} search with the all-sky coincidence search~\cite{Colaboration:2011np}.
This will allow us to estimate the sensitivity improvement offered by the targeted, coherent search.
For the all sky search, the background is one event per year at an \ac{SNR} of $10$ decreasing by two orders of magnitude per unit increase in \ac{SNR}.%
\footnote{This is taken from Figure 3 in \cite{Colaboration:2011np}, which shows a background of around $0.2$ events per year at \ac{SNR} of 10.  However, we must also apply a trials factor of six, as described in the paper, to give a background of 1 event per year at this \ac{SNR}.}
Interestingly, the background for the targeted, coherent search, as shown in \cref{fig:background_slides}, falls off at the same rate.
In both cases, this is significantly slower than expected in Gaussian noise, suggesting that both pipelines are affected in a similar way by the non-Gaussian transients in the data.
The background for the all-sky coincidence search translates to a \ac{FAP} of $10^{-3}$ in six seconds of data at an \ac{SNR} of 8.2.
In comparison, the targeted, coherent search achieves this background at an \ac{SNR} of 7.3.
While both of these are re-weighted \ac{SNR} measurements, and the details of the pipelines differ, the analysis methods have much in common, so it is reasonable to compare the results.
Thus, the coherent analysis provides approximately a 13\% reduction in the \ac{SNR} at a given \ac{FAP}.

We can use this to estimate the  benefit of performing the \ac{GRB} search.
To do so, we compare against a simple analysis that just examines the results of the all-sky search for triggers within the 6 second on-source window.
The comparison of \acp{FAP} above shows that the targeted, coherent search would identify a candidate event with a 13\% lower \ac{SNR}, or equivalently at a 13\% greater distance.
In addition, the targeted, coherent search applies lower single detector \ac{SNR} thresholds of 4, rather than 5.5, and it includes the \ac{SNR} contribution from all detectors, even if they did not produce a trigger above threshold.
For the case of \ac{GRB} 100928A, a signal near the detection threshold would be unlikely to register as a trigger in the Virgo detector, and the coherent analysis would register about 10\% greater \ac{SNR} by incorporating the power from Virgo.
This implies that the targeted, coherent search provides approximately a 23\% increase in distance sensitivity over a search that simply looks for a coincident \ac{GW} trigger from the all sky search.
This equates to approximately a doubling of the event rate.

\section{Search for Signals with Narrow Opening Angles}
\label{sec:face}
Short \acp{GRB} are believed to be beamed phenomena~\cite{Fong:2013lba,Panaitescu:2005er}, with prompt $\gamma\text{-ray}$ emission concentrated along collimated jets normal to the orbital plane.
These jets are expected to have opening angles of $<30 \degree$~\cite{Guelbenzu:2012id}.
Therefore, it may be reasonable to assume that observed short \ac{GRB} progenitor systems have their orbital angular momenta nearly parallel with the line-of-sight, corresponding to system orbital inclinations $\iota \sim 0$ or $\iota \sim \pi$ with respect to the observer.

\begin{figure}[tbp]
    \includegraphics[width=\linewidth]{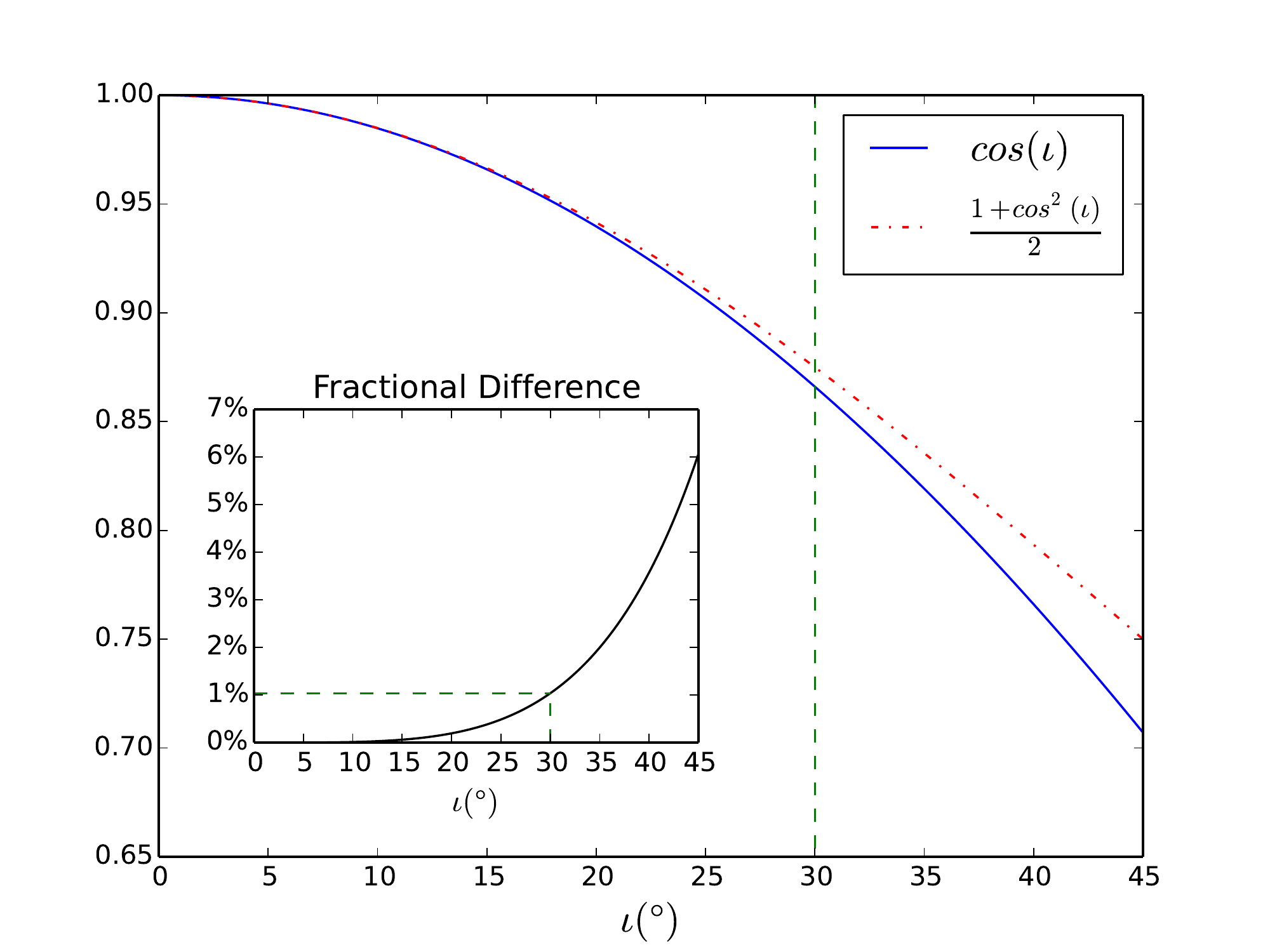}
    \centering
    \caption{Comparison between the $+, \times$ amplitude terms as a function of inclination angle $\iota$.             Note that even at $30\degree$ the difference is only $\sim 1$\%.
             \label{fig:inc_comp}}
\end{figure}

In \cref{eq:amps}, we see that the \ac{GW} amplitudes depend linearly on $\cos \iota$ and $(1+\cos^2 \iota)/2$.
For a binary inclination close to $\iota = 0$, both of these tend towards unity.
In \cref{fig:inc_comp}, we plot both amplitude factors as a function of $\iota$.
This serves to highlight the fact that the amplitudes vary almost identically with $\iota$, up to an angle of $30^{\circ}$, by which time they differ by only $\sim1\%$.
Even at $45^{\circ}$, the two amplitudes differ by only 6\%.
Consequently for \ac{GRB} signals, it is reasonable to treat the amplitude factors as equal and to approximate the signal as left circularly polarized.
Similarly, when $\iota \sim 180^{\circ}$, the two terms agree up to an overall sign and the signal is right circularly polarized.

It is therefore convenient to introduce a single amplitude and phase to describe the signal as
\begin{equation}
  \tilde{D} = \frac{D}{\cos \iota} \quad \mbox{and} \quad \chi_{l,r} =  \phi_0 \pm \psi \, .
\end{equation}
Then, for $\iota \approx 0$, the amplitudes simplify to
\begin{subequations}
    \begin{align}
	\mathcal{A}^1 \approx \mathcal{A}^4 &\approx -\frac{D_0}{\tilde{D}} \cos 2\chi_{l} \equiv \mathcal{B}_1 \, , \\
	\mathcal{A}^2 \approx - \mathcal{A}^3 &\approx \frac{D_0}{\tilde{D}} \sin 2\chi_{l} \equiv \mathcal{B}_2 \, , 
    \end{align}
\end{subequations}
and similar for $\iota \approx 180^{\circ}$.
As expected, the circularly polarized \ac{GW} signal is then dependent upon two amplitudes $\mathcal{B}_1$ and $\mathcal{B}_2$ (or, equivalently, a single overall amplitude and phase),
\begin{subequations}
    \begin{align}
	h_+(t) &= \mathcal{B}_1 h_0(t) - \mathcal{B}_2 h_{\pi/2}(t) \, , \\
	h_{\times}(t) &= \mathcal{B}_2 h_0(t) + \mathcal{B}_1 h_{\pi/2}(t) \, .
    \end{align}
\end{subequations}
rather than the original four amplitudes $\mathcal{A}^{\mu}$.

Substituting these expressions into \cref{eq:loglike}, and working in the dominant polarization, we obtain,
\begin{align}
    \ln\Lambda = & \mathcal{B}_1 \mathbf{(s|F_+ h_0 + F_\times h_{\pi/2})} 
    + \mathcal{B}_2\mathbf{(s|F_\times h_0 + F_+ h_{\pi/2})} \nonumber \\
    &- \frac{1}{2} \left[\mathcal{B}_1^2 + \mathcal{B}_2^2 \right]
    \left[\mathbf{(F_{+} h_{0} | F_{+} h_{0})} + \mathbf{(F_{+} h_{0} | F_{+} h_{0})} \right]
    \label{eq:loglikebs}
\end{align}
It is straightforward to maximize over the amplitude parameters $\mathcal{B}_{1,2}$ to obtain
\begin{equation}
    \label{eq:coh_face}
    \rho^2_{\mathrm{coh}} = \frac{\alpha^2 + \beta^2}{\mathbf{(F_{+} h_{0} | F_{+} h_{0})} + \mathbf{(F_{\times} h_{0} | F_{\times} h_{0})} } \, ,
\end{equation}
where
\begin{subequations}
    \begin{align}
	\alpha &= (\mathbf{s}|\mathbf{F}_+ \mathbf{h}_0) + (\mathbf{s}|\mathbf{F}_\times \mathbf{h}_{\pi/2}) \, , \\
	\beta &= (\mathbf{s}|\mathbf{F}_\times \mathbf{h}_0) - (\mathbf{s}|\mathbf{F}_+ \mathbf{h}_{\pi/2}) \, .
    \end{align}
\end{subequations}

The calculation proceeds in an analogous manner for $\iota \sim 180^{\circ}$, with the signal now right, rather than left, polarized.
After maximization, the coherent \ac{SNR} takes the same form as \cref{eq:coh_face}, but with
\begin{subequations}
    \begin{align}
	\alpha &= (\mathbf{s}|\mathbf{F}_+ \mathbf{h}_0) - (\mathbf{s}|\mathbf{F}_\times \mathbf{h}_{\pi/2}) \, , \\
	\beta &= (\mathbf{s}|\mathbf{F}_\times \mathbf{h}_0) + (\mathbf{s}|\mathbf{F}_+ \mathbf{h}_{\pi/2}) \, .
    \end{align}
\end{subequations}

The motivation for performing the search for only circularly polarized waveforms is to further reduce the noise background and thereby increase the sensitivity of the search.
Additionally, restricting to circularly polarized waveforms provides us with an additional \textit{null stream} that can be used to reject noise glitches.
Prior to assessing the improvement in real data, it is useful to evaluate the expected benefit in Gaussian noise.
The original search has four free amplitude parameters $\mathcal{A}^{\mu}$, and the coherent \ac{SNR} in the absence of a signal is $\chi^2$ distributed with four degrees of freedom.
When restricting to circular polarization, there are two free parameters $\mathcal{B}_{\mu}$ and the coherent \ac{SNR} in Gaussian noise will be $\chi^2$ distributed with two degrees of freedom.
However, we must now search over both left and right circularly polarized signals, which leads to a doubling of the number of trials.%
\footnote{The left and right circular waveforms are only orthogonal when the network is equally sensitive to both polarizations.  
For most sky locations, this is not the case, and the two trials are \textit{not} independent leading to a further reduction in the expected background.}
Comparison of these distributions, for a large number of trials, suggests restricting to circular polarization should result in at a decrease in \ac{FAP} of around one order of magnitude at fixed \ac{SNR}, or an increase in sensitivity at fixed \ac{FAP} of roughly 5\%.

\begin{figure}
    \centering
    \includegraphics[width=\linewidth]{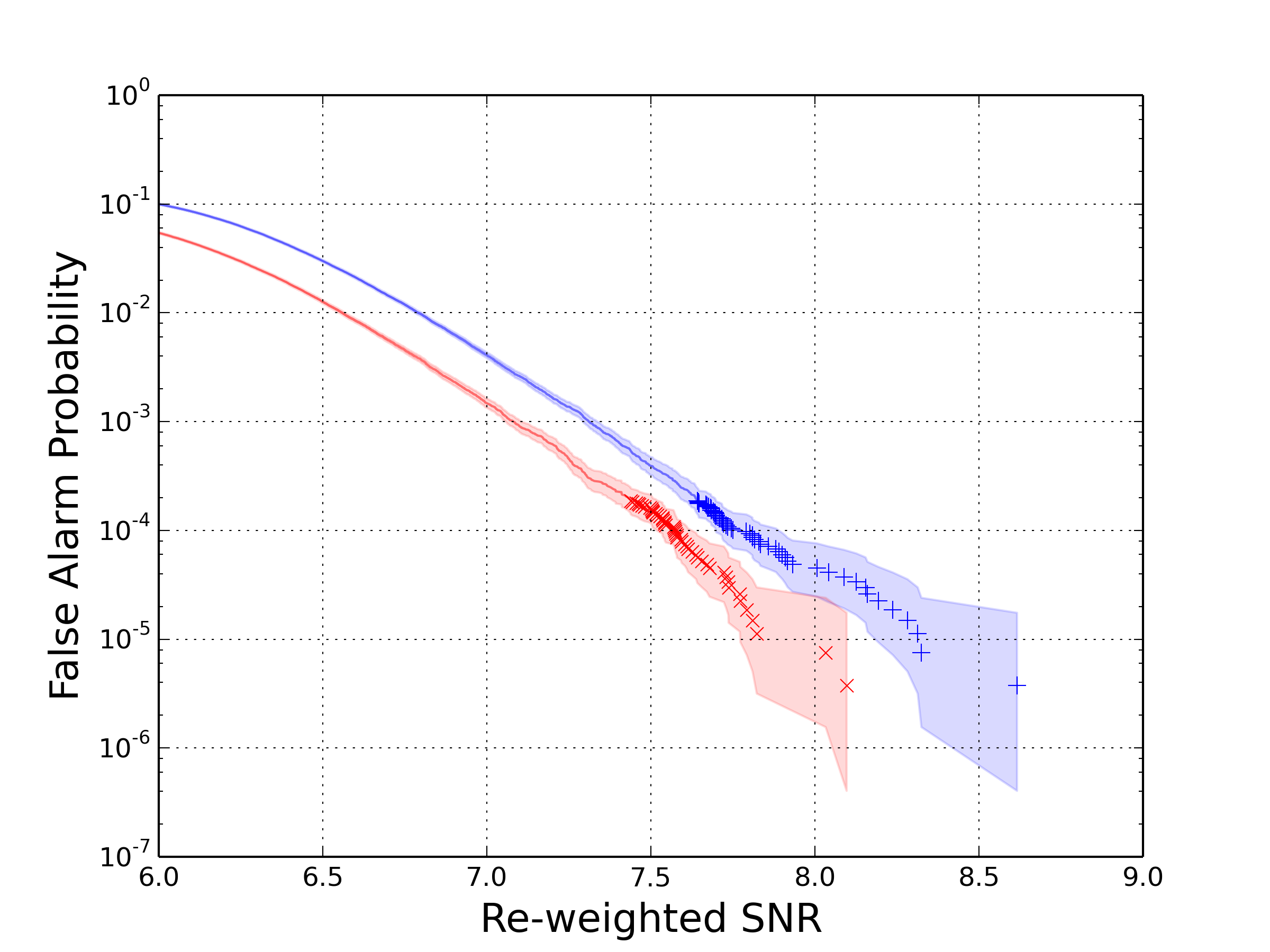}
    \caption{The background significance against detection statistic for a search performed for \ac{GRB} 100928A.
             In red $\times$, we plot the background calculated using the circular polarization restriction and in blue + we plot the background from the un-restricted search.
             In both cases, we perform time shifts of the data as discussed in \cref{sec:back}.
             Over a broad range of \ac{SNR} values, the circular polarization restriction reduces the background by a factor of three.
             Equivalently, the required \ac{SNR} to achieve a given \ac{FAP} is reduced by about 0.25, equating to a 3\% increase in the distance sensitivity of the search.
             For clarity of presentation we have only plotted the loudest 50 trials for each search.
    \label{fig:background_fo}
    }
\end{figure}

In \cref{fig:background_fo} we plot the \ac{FAP} as a function of \ac{SNR} for the circularly polarized and un-restricted searches.
Over a broad range of \acp{SNR} we observe a reduction in the background of a factor of three, corresponding to an increase in sensitivity of around 3\% at a given \ac{FAP}.
This improvement is less significant than might have been expected in Gaussian data, and only serves to emphasize that the data we are using is not Gaussian, and events in the tail of the distribution are due to noise transients in the data.

Interestingly, we have noticed that the most significant background triggers in the circular search do not correspond to outliers in the un-restricted search.
This is likely due to how the pipeline selects triggers.
It first applies a clustering method to choose the trigger with the largest coherent \ac{SNR} in a given time window, before applying signal consistency tests to the trigger which may lead to it being discarded or the \ac{SNR} re-weighted.
Consequently, it is possible that loud events in the un-restricted search do not survive in the circular analysis, and vice versa.

We have demonstrated that restricting to circularly polarized signals can provide a small improvement in the search sensitivity and, furthermore, that it is a reasonable approximation given our current understanding of \ac{GRB} beaming.
We note that a 3\% improvement in distance reach corresponds to a 10\% increase in the rate of observable signals.

\section{Searching a Patch of the Sky}
\label{sec:sky}
Short \acp{GRB} are localized to sky error boxes of varying sizes by different satellites.
This has implications for the targeted \ac{GW} search following up on these events.
For example, the \ac{BAT} instrument aboard NASA's \textit{Swift} satellite is capable of localizing to 1-4 arcminutes~\cite{Barthelmy:2005hs}, while the typical \ac{GW} localization region is several square degrees or larger~\cite{Fairhurst:2009tc, Aasi:2013wya}.
Thus, we may follow up a \ac{BAT} trigger by searching only a single point on the sky since the \ac{GRB} localization is significantly better than the sky resolution of the \ac{GW} search.
However, the \ac{GBM} aboard NASA's \textit{Fermi} satellite often localizes \acp{GRB} to far larger patches of the sky~\cite{Meegan:2009qu}.
The $3 \sigma$ confidence regions are roughly circular, with a radius of several degrees.
Additionally, the \ac{IPN} localizes \acp{GRB} by triangulation with a number of satellites~\cite{Hurley:2002wv}.
Depending upon the number of satellites observing the event and their relative positions, the localizations can range from under a square degree to hundreds or even thousands of square degrees.
For poorly localized short \acp{GRB} observed by \textit{Fermi} or \ac{IPN}, the \ac{GRB} localization will be comparable to, or larger than, the typical \ac{GW} localization region.
Consequently, it is no longer appropriate to treat the \ac{GRB} localization as a single point in the sky, and we must extend the \ac{GW} search to cover the entire confidence region.

The targeted, coherent \ac{GW} search makes use of the sky location in two ways.
Firstly, and most importantly, it is the sky location which determines the relative arrival time of a signal at the detectors in the network.
These time delays are used to appropriately shift the data prior to coherently combining them in the search.
Using the incorrect sky location will cause the signals from different detectors to be mis-aligned in time.
Secondly, the detector sensitivities, encoded in the antenna response factors $F_{\{+,\times\}}$, depend upon the location of the source relative to the detector.
The use of incorrect $F_{\{+,\times\}}$ will lead to the wrong weighting of detector data streams in the coherent \ac{SNR} and signal power being present in the null stream.

We can estimate when the single sky point search will not be sufficient.
To do so, let us consider only the loss in \ac{SNR} arising from timing offsets.
In a matched-filter search, the recovered \ac{SNR} in a detector falls off as
\begin{equation}
    \rho(dt)^{2} \approx \rho_{o}^2 [1 - (2\pi \sigma_{f})]^{2} dt^{2} \, ,
    \label{eq:del_snr}
\end{equation}
where $\sigma_{f}$ is the signal bandwidth, which is typically around 100 Hz for a binary merger signal~\cite{Fairhurst:2009tc}.
Thus, a timing offset of $\delta t = 0.5\text{ms}$ will lead to a 5\% loss in \ac{SNR} in a single detector.

Given a network of $N$ detectors, $D_{\{1,\ldots,N\}}$, let $\mathbf{r}_i$ denote the location of the detector and $t_i$ be the arrival time of the \ac{GW} signal at detector $i$ from a \ac{GRB} at the central location of the sky patch.
The distance between two detectors is
\begin{equation}
    d_{ij} = \vert\vert \mathbf{r_j}-\mathbf{r_i}\vert\vert \, ,
\end{equation}
and the light travel time between them is
\begin{equation}
    \label{eq:ltt}
    T_{ij} = d_{ij}/c \, .
\end{equation}
The difference in the arrival time of the signal at two detectors, $\tau_{ij} $, is calculated as~\cite{Rabaste:2009mx},
\begin{equation}
    \tau_{ij} = t_i - t_j = \frac{1}{c} \left(\mathbf{r_i}-\mathbf{r_j}\right) \cdot \mathbf{w}  \equiv T_{ij} \cos\alpha \, ,
\end{equation}
where $\mathbf{w}$ is the unit wave vector describing the direction of propagation of the source, and $\alpha$ is the angle between the line connecting the detectors and the direction to the source.

It is then straightforward to calculate the change in time delay with a change in the angle $\alpha$ as
\begin{equation}
  \delta \tau_{ij} = \sqrt{ T_{ij}^{2} - \tau_{ij}^{2} }\, \delta \alpha \, .
\end{equation}
So, for a source lying on the line connecting the two detectors, the time delay $\tau_{ij}$ between detectors is maximal and changes only quadratically with the change in the location of the source.
In contrast, for a source which lies on the zero time delay plane, $\tau_{ij} = 0$, a change in location will induce the largest time offset.

Once we select the maximum time offset $\delta t$ that we are willing to tolerate, it is straightforward to calculate the required angular spacing of the sky points as
\begin{equation}
    \delta\alpha = \min_{i,j} \left[\frac{2 \delta t}{\sqrt{T_{ij}^2 - \tau_{ij}^2}}\right] \, .
\end{equation}
Here, the factor of two arises because $\delta t$ is the largest single detector time offset.
We typically choose $\delta t = 0.5\text{ms}$.
The two \ac{LIGO} detectors are separated by a light travel time of 10ms, while LIGO and Virgo are separated by around 25ms, which sets the angular scale to around $2\degree$ for the LIGO detectors and $1\degree$ between LIGO and Virgo.
In practice, the resolution is usually determined by the detector pair $(D_i,D_j)$ for which the \ac{GRB} target location has smallest relative arrival time difference.

\begin{figure}[tbp]
    \centering
    \includegraphics[width=\linewidth]{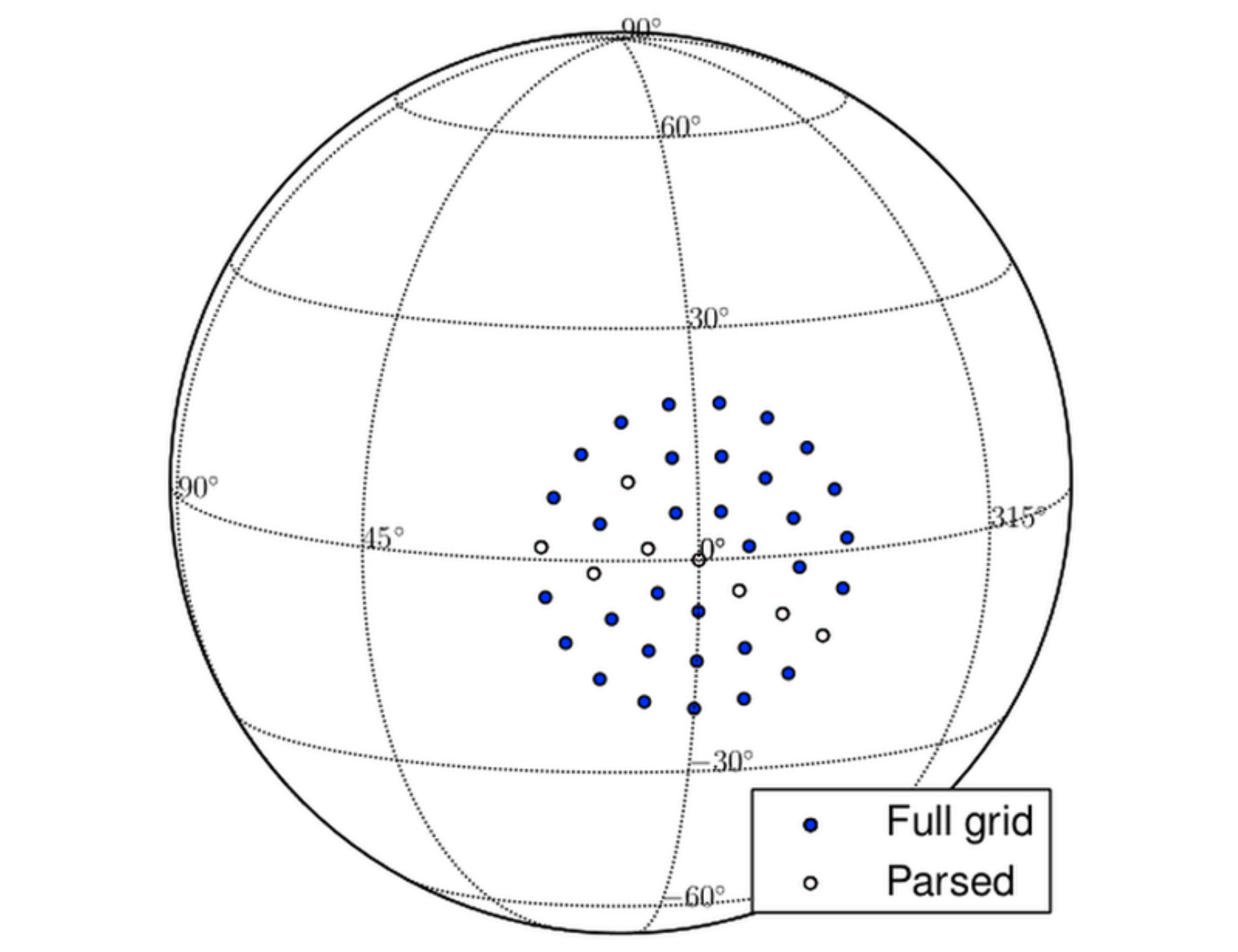}
    \caption{An example patch of sky points projected onto the celestial sphere.
             The blue filled circles show the full grid, while the empty circles are those few points that map to unique differences in signal arrival time between \ac{LIGO}'s Hanford and Livingston detectors.
             The parsed points do not form a straight line, but this is simply due to an artifact of the parsing routine and has no effect on the grid reduction.}
    \label{fig:two_site_circular_grid}
\end{figure}

The circular grid is generated by placing rings of points spaced by $\delta\alpha$, starting at the center, with the final ring passing the $3 \sigma$ confidence radius.
An example of such a grid is shown in \cref{fig:two_site_circular_grid} (full grid).
Each ring will have $2\pi n/ \delta\alpha$ points, where $n=0$ labels the central point and increases as we move outwards.  The method of covering the patch is based upon the one introduced in \cite{Aasi:2014ent}.
In the analysis, each point in the grid is treated independently, with the single-detector data streams time shifted appropriately for the given sky location.
The coherent \ac{SNR} and signal consistency tests are calculated with the appropriate detector responses, $F_+$ and $F_\times$, for that sky point.
As with the background estimation, searching over points in the sky patch is performed \textit{after} the computationally dominant step of calculating the single detector \ac{SNR} time series.
Consequently, \acp{GRB} observed by \textit{Fermi} GBM, requiring around hundred sky points, are processed in approximately double the time required for the \textit{Swift} \acp{GRB} with a single sky point.

\begin{figure}[tbp]
    \centering
    \includegraphics[width=0.5\textwidth]{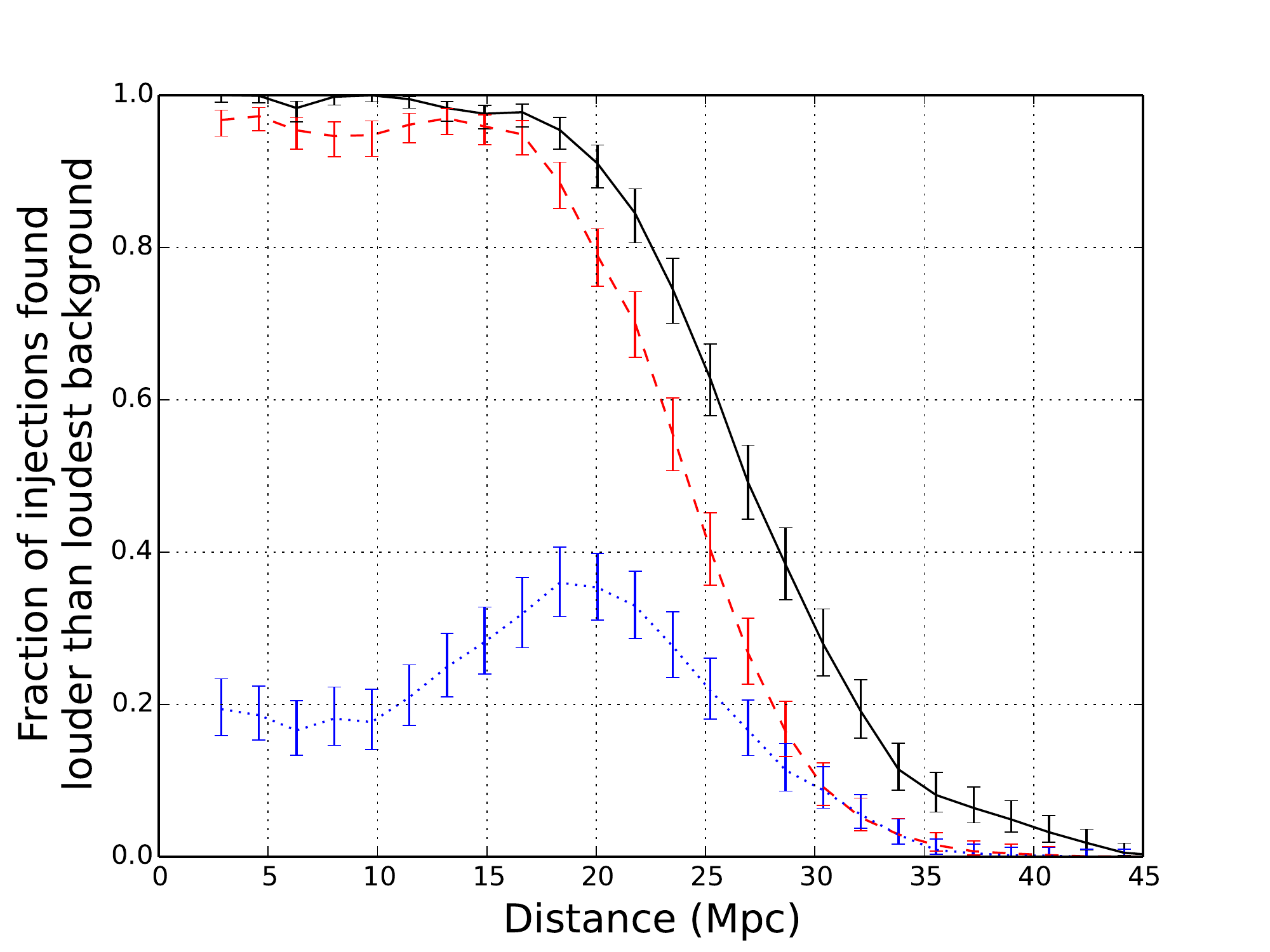}
    \caption{The fraction of artificially injected binary neutron star signals found louder than the loudest background event as a function of injected distance.
             The three curves represent three observational scenarios for a three detector network comprised of Virgo and both \ac{LIGO} interferometers.
             In the scenario mimicking a \ac{BAT} \ac{GRB} (black solid line, error radius $= 0.036\degree$) the pipeline searches a single point on the sky and finds 90\% of signals within 20 Mpc.
             In the two scenarios mimicking a \ac{GBM} \ac{GRB} we see that by searching over a patch of points covering the large error box of $15\degree$ radius (red dashed line) the pipeline performs nearly as well as for the \ac{BAT} \ac{GRB} for signals below 15 Mpc.
             This is in stark contrast to the previous treatment for \ac{GBM}-like \acp{GRB} (blue dotted line), which searched a single point at the center of the error box resulting in very poor rates of injection recovery.
             The increased number of trials resulting from multiple sky points leads to a tail of background events louder than any seen of the \ac{BAT} single point search, reducing the overall sensitivity of the patch search.}
    \label{fig:eff_comp}
\end{figure}

To demonstrate the efficacy of searching over a sky patch, we repeated the analysis of \ac{GRB} 100928A, but used a typical \textit{Fermi} GBM 3$\sigma$ localization uncertainty radius of $15\degree$~\cite{Meegan:2009qu}.
The sky patch for the search contained 178 search points in total.
When performing simulations, the location of each source was chosen randomly from a normal distribution with width $5\degree$, ie. $\sim99\%$ of simulated signals were within the $15\degree$ radius $3\sigma$ localization region.
In the rest of this section, we use the results for \ac{BNS} systems exclusively to illustrate the effect of a search over a patch of the sky.
However, similar effects are observed in searches for \ac{NSBH} systems.

In \cref{fig:eff_comp}, we show the search efficiency as a function of distance for three different searches: a single point search with simulations spread over the $0.036\degree$ \textit{Swift} \ac{BAT} sky patch; a single point search with simulations spread over a typical $15\degree$ \textit{Fermi} \ac{GBM} sky patch; and a grid of points covering the \ac{GBM} sky patch with simulations spread over the patch.
In all cases the efficiency is calculated at the \ac{SNR} of the loudest background event in the \textit{short slide} analysis.
If we perform the search using only a point at the center of the \textit{Fermi} localization region, the results are poor: across the whole range of distances, the search efficiency is never greater than 40\%, even for nearby signals which have large \acp{SNR}.
The reason for this lies in the signal consistency tests discussed in \cref{sec:summary}.
At the incorrect sky location, the signal does not match the template due to inevitable time offsets between them and the signal will be recovered with a different phase in each of the detectors.
Consequently, the coherent \ac{SNR} will not correctly reflect the total signal power and this will lead to increased values of the signal consistency tests, and in particular lead to a significant amount of power in the null \ac{SNR}.
Thus, many signals at smaller distances are found with large total \ac{SNR} values but also have sufficient power in the null \ac{SNR} statistic to fail the hard cut in \cref{eq:null_cut} and are therefore rejected as potential detections.

The sensitivity of the search over the \textit{Fermi} error region is almost the same as the search over just the \textit{Swift} point at small distances, but decreases more rapidly for quieter signals at larger distances.
For example, the distance at which we achieve 50\% efficiency is reduced by 10\%.
This loss in sensitivity can be attributed to the fact that we have searched over numerous sky points thereby increasing the number of background trials.
In this example, we obtain a loudest background event with re-weighted \ac{SNR} value of 8.33 compared to 7.51 for a single point, which is consistent with the loss of reach of the search.

This method of placing a grid of points in the sky has already been used in the analysis of \textit{Fermi}-detected \acp{GRB} during LIGO Science Run 6 and Virgo Science Runs 2 \& 3.
An analogous method was used to perform the search over the irregular sky patches produced by the \ac{IPN}~\cite{Predoi:2011aa}.

\subsection{Two-site time-delay degeneracy}

\begin{figure}[tbp]
    \centering
    \includegraphics[width=0.5\textwidth]{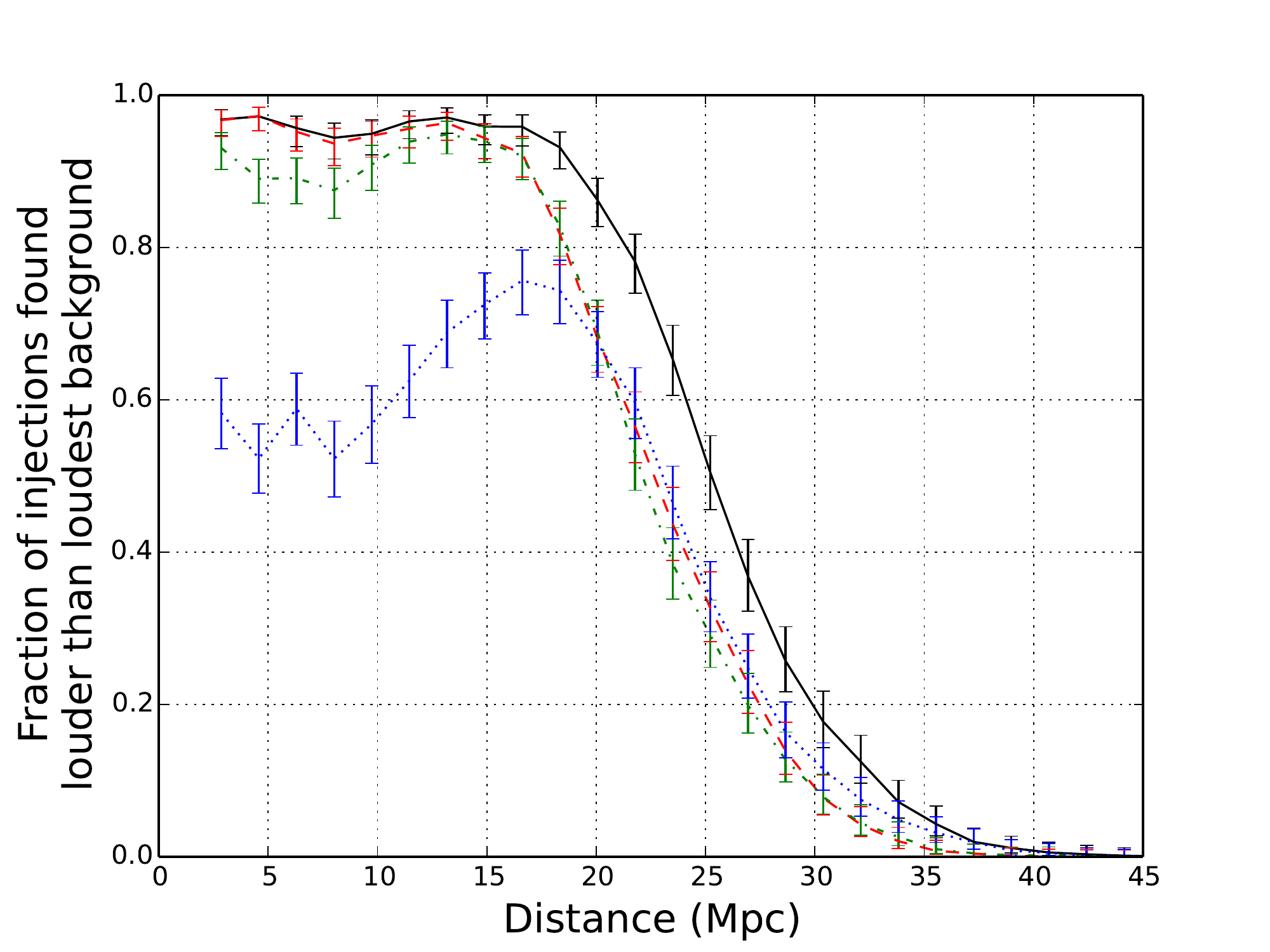}
    \caption{The fraction of artificially injected binary neutron star signals found louder than the loudest background event using only the \ac{LIGO} Hanford and Livingston detectors, plotted as a function of injected distance.
             As in \cref{fig:eff_comp}, we plot a scenario mimicking a \ac{BAT} \ac{GRB} (black solid line, error radius $= 0.036\degree$) where the pipeline searches a single point on the sky.
             In this case, the pipeline finds 90\% of signals within 18 Mpc.
             In the scenario where a \ac{GBM} \ac{GRB} with error box of $15\degree$ radius is searched at a single point (blue dotted line), we see poor signal recovery performance at small distances due to signal consistency effects, similar to the three detector case.
             The difference between the full patch of search points (red dashed line) and a set of points covering unique time delays between sites (green dot-dashed line) is noticeable at small distances, with the use of incorrect antenna response factors causing a drop in performance for the parsed patch.
             Again, the increased number of trials resulting from multiple sky points leads to a tail of background events louder than any seen of the \ac{BAT} single point search, reducing the overall sensitivity of multiple point searches.}
    \label{fig:eff_HL}
\end{figure}

In the case of a two-site detector network, for example the \ac{LIGO}-only network, the ability to resolve independent sky locations is vastly reduced.
With a single baseline between sites, multiple sky locations will map to the same difference in signal arrival time.
Thus, when moving across the sky patch, there will be one direction where only the antenna response factors $F_{\{+,\times\}}$ change, and not the time delays, while in the orthogonal direction both will change.
With two detectors, after maximizsing over the $\mathcal{A}^{\mu}$, the values of $F_{\{+,\times\}}$ drop out of the coherent \ac{SNR} expression.
This is not immediately obvious, but can be understood by noting that for a two detector search, there are four degrees of freedom in both the coincident and coherent searches.
Therefore, \textit{any} observed amplitude and phase in the two detectors is consistent with a astrophysical signal; there is no null stream.
Then, the size of the sky grids can be significantly reduced, to represent only those sky locations that map to unique time-delays between observatory sites.
\cref{fig:two_site_circular_grid} shows an example result of parsing the circular sky maps to remove degeneracies in time-delay.
For the map shown, only 20\% of the points are required to uniquely span the allowed time-delays between the \ac{LIGO} sites, allowing a reduction in cost in the analysis for two-site \ac{GRB} analyses.

Unfortunately, once we restrict to circularly polarized signals, as described in \cref{sec:face}, the restriction to a single time-delay line is no longer appropriate.
Now, there are only two free signal amplitudes, which cannot match arbitrary amplitude and phase measurements in the two detectors.
Thus the detector response functions are again enter into the construction of the coherent \ac{SNR} and the circular null stream.

In \cref{fig:eff_HL}, we show the sensitivity of the search performed using only the two \ac{LIGO} detectors in Hanford and Livingston and incorporating an inclination restriction.
As before, we plot the \textit{Swift} search results -- where both the simulated signals and search are restricted to a single sky point -- as a reference.
Next we consider the \ac{GRB} localized to a typical \textit{Fermi} \ac{GBM} error region.
When searching over the full \textit{Fermi} sky patch, there is again a degradation of the sensitivity due to a tail of loud background events (a maximum \ac{SNR} of 8.12 compared to 7.25 for the single point search).
However, searching of a single sky point leads to a dramatic loss of sensitivity, with only 60\% of nearby signals being recovered.
By searching over only the one dimensional time-delay space, we recover the majority of this sensitivity, but do observe a small drop in efficiency at low distances due to the use of incorrect antenna response factors.

\section{Discussion}
\label{sec:discuss}
The advanced \ac{LIGO} and Virgo detectors will be sensitive to binary merger signals from hundreds of Mpc, or even Gpc in the case of \ac{NSBH} systems.
These distances are then comparable to those of the closest \acp{GRB} and a joint observation in the coming years is a distinct possibility.
In this paper, we have presented details of an improved GW--GRB search that implements several new features that will facilitate joint observations.
The work extends that in~\cite{\HF} in three distinct ways.
First, we have introduced a method of time-shifting the background data in order to estimate false alarm probabilities lower than $10^{-5}$.
An event of this significance, or greater, will likely be required to claim the first joint GW--GRB observation.
Critically, we have seen that there is no ``tail'' of rare, high \ac{SNR} events that would hinder a detection claim.
Second, we have developed a method for searching over extended regions of the sky, rather than just a single point.
The majority of short \ac{GRB} observations are currently made with the \textit{Fermi} GBM detector, which typically localizes events to tens of square degrees.
With the capability of searching sky patches, we can now achieve a comparable sensitivity for \textit{Fermi} \ac{GBM} bursts as to those which are localized to arc-second accuracy with \textit{Swift}.
Third, we have made use of astrophysical priors on \ac{GRB} jets to restrict the search to nearly face-on binaries whose gravitational wave signal will be circularly polarized.
We have shown that this provides a small, but significant, improvement in sensitivity of the search.
Taken together, these improvements to the search mean it is ready to be deployed in the future on advanced \ac{GW} detector data at the time of short \acp{GRB}.
Nonetheless, there are several additional features that we plan to implement in the near future, which we describe below.

We would like to provide rapid, \ac{GCN} style alerts of the \ac{GW} search results for \acp{GRB}.
For these to be useful, the analysis must be completed as rapidly as possible.
This can be achieved by a simple re-ordering of the analysis to prioritize the on-source analysis with short time slides only to provide an initial result, with a \ac{FAP} measured to $10^{-4}$ within an hour or two of the \ac{GRB}.
Subsequent improvements on \ac{FAP} measurement and search sensitivity using simulated signals will follow later if a promising \ac{GW} candidate is found.
To make this feasible in the long term, a process is being developed to automatically launch the analysis upon receipt of a \ac{GCN} alert.
 
The current search makes use of template waveforms appropriate for binaries with non-spinning components.
For neutron stars, this is a reasonable approximation as they are expected to have low spins which will not greatly affect the waveform~\cite{Brown:2012qf}.
However, in an \ac{NSBH} system, the black hole spin can have a significant effect on the emitted waveform.
The component of the spin aligned with the orbital angular momentum will affect the rate at which the binary inspirals~\cite{Schmidt:2012rh}, while the orthogonal spin components will lead to precession of the system~\cite{Apostolatos:1994mx}.
It has been shown~\cite{Ajith:2011ec,Harry:2013tca} that using waveforms which incorporate the effects of aligned spins can greatly enhance the sensitivity of a search to \ac{NSBH} systems.
Furthermore, when the spin is aligned with the orbital angular momentum, the waveforms simplify to the form given in \cref{eq:ht}.
Thus, it is straightforward to simply extend the template bank to include these waveforms and incorporate the effects of aligned spins.
Including the aligned spin contribution to the waveform will aid the sensitivity of the search.
It is not so straightforward to incorporate precession effects.
Thankfully, precession typically has a less significant effect on the waveform when the binary is observed close to face on~\cite{Hannam:2013oca}, so that will reduce the importance of precession for the \ac{GRB} search.
Nonetheless, we would like to incorporate these effects.
In~\cite{Harry:2011qh}, we investigated a method of extending the search to waveforms with precession.
In the future, we will identify the regions of parameter space where the spin-aligned waveforms do not provide good sensitivity to precessing signals and complete the development of the analysis in~\cite{Harry:2011qh} to provide a sensitive search over these parts of the parameter space.

Not all \ac{NSBH} mergers are expected to emit electromagnetically~\cite{Foucart:2012nc,Stone:2012tr,Palenzuela:2013hu}.
Under most scenarios, electromagnetic emission requires the formation of a torus around the central black hole.
In~\cite{Pannarale:2014rea}, the region of black hole masses and spins which might give rise to this torus was investigated.
Around half of the \ac{NSBH} parameter space will not lead to torus formation, under any reasonable model of a neutron star equation of state, and can therefore be eliminated from the analysis.
By eliminating these templates from the analysis, we can reduce both the computational cost and search background.
This is a feature we plan to implement in the near future.

The advanced detectors are sensitive to signals from 10 Hz upwards~\cite{Harry:2010zz}, in comparison to 40 Hz for the initial detectors.
A \ac{BNS} system will take 1,000 seconds to evolve from 10 Hz to merger.
Consequently, the search must be extended to deal with longer duration templates, in order to capture all of the available power in the signal.
This can be achieved by extending the lengths of the analyzed data segments in the search, although significant changes will be required to handle 1,000 second templates.
However, the early advanced detector runs are not expected to obtain the full low-frequency sensitivity~\cite{Aasi:2013wya}.
For these, a search beginning at 25 Hz (templates of 90 second duration) would capture the vast majority of the signal power.
This can be achieved relatively easily, and will be available for the initial runs that are expected in 2015.

\begin{acknowledgments}
The authors would like to thank the LIGO Scientific Collaboration and the Virgo Collaboration for allowing access to the data used in generating the results in \cref{sec:back,sec:face,sec:sky}.
In addition, they thank Patrick Sutton for recommending 100928A as a suitable example GRB, and Dipongkar Talukder for comments on the manuscript.
This work was supported by the UK Science and Technologies Funding Council through grant ST/L000962/1, the Royal Society and the US National Science Foundation through grants PHY-0847611, PHY-1205835, PHY-1404395, and AST-1333142.
\end{acknowledgments}

\bibliographystyle{apsrev4-1}
\bibliography{}

\end{document}